\documentclass[12pt]{spieman}  

\usepackage{amsmath,amsfonts,amssymb}
\usepackage{graphicx}
\usepackage[colorlinks=true, allcolors=blue]{hyperref}
\usepackage{float}
\restylefloat{table}
\usepackage[section]{placeins}
\usepackage[export]{adjustbox}
\usepackage{soul}
\usepackage{booktabs}

\title{Performance analysis of a Hadamard Transform Spectral Imaging system}

\author[a]{John Nijim}
\author[a]{Zoran Ninkov}
\author[b]{Dmitry Vorobiev}
\author[a,c]{Kevin Kearney}
\affil[a]{Rochester Institute of Technology, One Lomb Memorial Dr., Rochester, NY 14623-5603}
\affil[b]{Laboratory for Atmospheric and Space Physics, 1234 Innovation Drive, Boulder, CO, 80303}
\affil[c]{Starris Optimax Space Systems, 6367 Dean Pkwy, Ontario, NY 14519}
 
\begin{document} 
\maketitle

\begin{abstract}
Hadamard Transform Spectral Imaging (HTSI) is a multiplexing technique used to recover spectra via encoding with multi-slit masks, and is particularly useful in low photon flux applications where signal-independent noise is the dominant noise source. This work focuses on the procedure that is used to recover spectra encoded with multi-slit masks generated from a Hadamard matrix; the decoding process involves multiplying the output encoded spectral images by the inverse of the Hadamard matrix, which separates any spectra that were overlapping in the target object. The output from HTSI is compared to direct measurement methods, such as single-slit scanning, to evaluate its performance and identify under which conditions it can provide an advantage or disadvantage. HTSI resulted in an increase in the average signal-to-noise (SNR) ratio of spectra when signal-independent noise, such as detector read noise, is present, and has no average net effect when signal dependent-noise, such as Poisson photon noise, is the only noise source present. The SNR of emission lines  was found to be greater with HTSI than with single-slit scanning under both signal-independent and signal-dependent noise, and increases as the ratio of read-to-shot noise increases.  
\end{abstract}


\section{Introduction}
\label{sec:intro}  

A spectrometer (or spectrophotometer) is a device used to measure the wavelength spectrum of light reflected or emitted from a sample. Typically, the spectrometer is focused on a small region of the sample, collecting spectra from that area. This approach prevents the overlap of spectra from different regions of an extended source, which could otherwise result in uninterpretable data. If spectra from every point on the object are desired, an imaging spectrometer is required. The spectral data collected from an extended source is often referred to as a ``datacube'', reflecting its three-dimensional nature, with two axes corresponding to spatial dimensions (vertical and horizontal) and the third representing wavelength. Imaging spectroscopy is widely used across various scientific fields, including astronomy, biology, environmental monitoring, and materials science.

A simple imaging spectroscopy method is to scan the focused spot over the entire region – collecting the spectrum at each location, while another involves imaging the region onto a slit and then scanning the slit across the area. While spot- or slit-scanning can provide a solution, it is inefficient for extended sources. Further, many sources are characterized by weak signals that often result in degraded spectral resolution, particularly in applications such as species identification or the analysis of spectral line widths and ratios. To overcome these limitations, spatial multiplexing methods can be used. These methods allow multiple regions of the source to be observed at once, enhancing sensitivity and efficiency. One such multiplexing technique uses a series of coded-aperture masks to replace the slit, where each mask allows 50\% of the light from the source through the system. While it is true that the spectra from different regions now overlap, the spatial encoding process allows the unscrambled datacube to be mathematically recovered. Harwit and Sloane (1979) \cite{Harwit_Sloane_1979} demonstrated that Hadamard matrices provide an optimal encoding for multiplexed signals, enabling high information recovery. The spectra are projected along the Hadamard basis vectors (rows or columns of a Hadamard matrix), which form a complete orthogonal set. Using this Hadamard Transform Spectral Imaging (HTSI) process, the signals can be reconstructed using an inverse transform, which acts as a linear unbiased estimator, even in the presence of noise. This method significantly enhances the resolution and fidelity of spectral data, making it particularly useful in applications that demand high spectral precision.

\begin{figure}[H]
    \centering
    \includegraphics[width=1\linewidth]{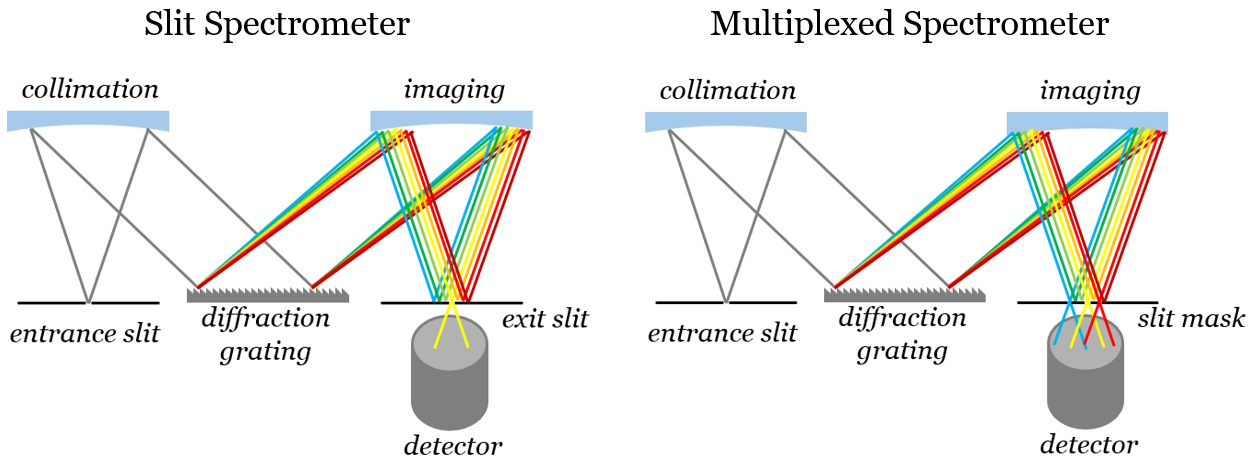}
    \caption{\label{fig:htsi_sketch}In its most simple application, Hadamard transform spectroscopy can be performed by replacing the exit slit of a monochromator with a series of slit masks. Thus, the spectral image is sampled by cycling through the slit masks, rather than by scanning the spectral image along the exit slit. This is very similar to the concept of a multiplexed single-pixel camera. See Harwit and Sloane (1979)\cite{Harwit_Sloane_1979} for an excellent review.}
\end{figure}

The HTSI method is of most use when the signal is detector-noise limited (low signal), as opposed to photon noise (Poisson) limited.  As demonstrated by Streeter et al., \cite{Streeter_Burling-Claridge_Cree_Künnemeyer_2009} HTSI results in no degradative effect - on average - on the noise level due to Poisson noise sources while preferentially reducing additive (detector) noise.  The result at low signal levels is a SNR boost proportional to the square root of the number of masks $n$ used in the acquisition, converging to a boost of unity at high signal levels. Per Streeter, ``the SNR boost is delivered when most needed.''

\subsection{HTSI implementation}
Hadamard Transform Spectral Imaging is an extension of Hadamard spectroscopy, which was originally developed to improve the signal-to-noise of measurements performed with a conventional spectrometer. The monochromator slit was replaced with a multi-slit mask (Fig. \ref{fig:htsi_sketch}), with the slit size and position determined by elements of a Hadamard matrix. In this configuration, a single-element detector either observes the light narrowly filtered by the exit slit or by one of several slit masks. The slit masks are typically made of opaque materials, designed to block 50\% of the spectral image at a time. The exact arrangement of the slits in each mask is determined by elements in a row of a Hadamard matrix of rank $n$. This configuration (modulation taking place in the image plane, in this case of a spectral image) is similar to that of a single-pixel camera. 

The Hadamard matrix can be constructed with one of several methods, with all Hadamard matrices required to have certain properties, including: 1) The elements of a Hadamard matrix are +1 or -1 and, except for the first row, each row has an equal number of +1 and -1 elements and 2) any two distinct rows of a Hadamard matrix are orthogonal (their scalar product is zero). Hadamard matrices can be constructed using certain algorithms, such as Sylvester's construction (Eq. \ref{eq:h_gen}); however, many more Hadamard matrices exist than ones which can be constructed with any specific algorithm.

\begin{equation}
\label{eq:h_gen}
H_{2n} = 
\begin{bmatrix}
H_n & H_n\\
H_n & -H_n
\end{bmatrix}
= H_2 \otimes H_n
\end{equation}

\begin{equation}
\label{eq:h4}
H_4 = 
\begin{bmatrix}
1 & 1 & 1 & 1\\
1 & -1 & 1 & -1\\
1 & 1 & -1 & -1\\
1 & -1 & -1 & 1
\end{bmatrix}
\end{equation}

\begin{figure}[h!]
\begin{center}
\includegraphics[width=0.7\linewidth]{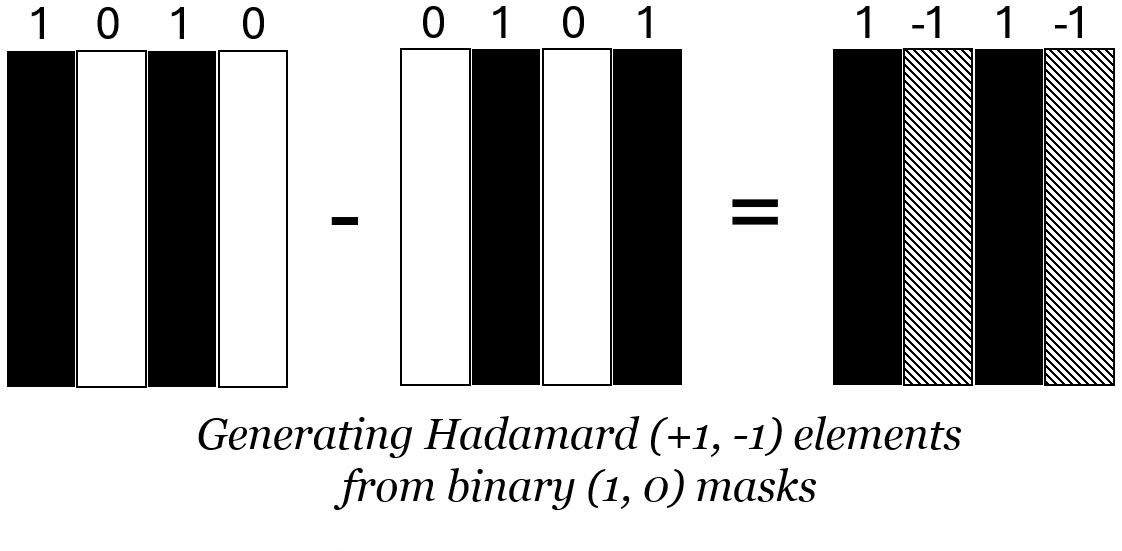}
\end{center}
    \caption{\label{fig:elements}In HTSI, rows of a Hadamard matrix are used to generate ``bar code'' style slit masks to multiplex the spectral image; however, the negative elements cannot be created directly, each row must be decomposed into two complimentary binary masks, whose difference recreates a Hadamard row. Here, an example for the 2nd row of a rank 4 Hadamard matrix (Eq.\ref{eq:h4}) is shown. This procedure is repeated for all rows of the matrix. Ideally, an HTSI spectrograph can acquire the images multiplexed by the binary masks simultaneously.}
\end{figure}

The elements of a Hadamard matrix cannot be used directly to generate slit masks because the negative elements require negative intensity of light. As such, in practice the Hadamard elements (+1, -1) are decomposed into a difference of two patterns of binary elements (1, 0) (Fig. \ref{fig:elements}). Two matrices can be created from a Hadamard matrix, one for the positive elements and the other for the negative via the following relations: $H^+=(1+H)/2$ and $H^-=(1-H)/2$\cite{Streeter_Burling-Claridge_Cree_Künnemeyer_2009}. Here, \begin{math}H\end{math} is a Hadamard matrix, and $H^+$and $H^-$ are the matrices containing binary elements (1s and 0s) representing the positive and negative elements, respectively. The difference of these two matrices reconstructs the original Hadamard elements. Traditionally, 1s map to transmissive regions of the mask, and 0s map to regions that block light.

A Hadamard transform of the data to be measured can then be done opto-mechanically, by placing each mask one at a time in the optical path, and recording the measurement. Each mask corresponds to a multiplexed observation which projects the signal along the Hadamard basis vectors. Let $\eta$ be the vector containing all multiplexed observations. If the signal to be measured is $\psi$, then the multiplexed observations are $\eta = H_n \psi$. To recover $\psi$, the inverse of the transform must be computed, which can be achieved using the property that Hadamard matrices satisfy (Eq. \ref{eq:h_prop}):

\begin{equation}
\label{eq:h_prop}
H_n H_n^T = H_n^T H_n = n I_n
\end{equation}

\noindent This implies that $H_n^{-1} = H_n^T / n$, and therefore $\psi = (1/n)H_n^T \eta$.\cite{Harwit_Sloane_1979}

In practice, these observations can be acquired with spectrograph instruments that have one or two spectral channels. A dual-channel spectrograph will obviously have the advantage that both the ``positive" and ``negative" signal observations can be taken simultaneously. A single detector can still preform HTSI by taking each observation one at a time in sequence, however this would double the observation time. Examples of such instruments are described in Section 4. If a short observation time is desired with a single-channel spectrograph, an alternate matrix whose elements are already 1s and 0s can be used, but provides a somewhat lower SNR increase \cite{Oram_Ninkov_2020}. Harwit and Sloane\cite{Harwit_Sloane_1979} introduce simplex matrices (\textit{S}-matrices) as the optimal encoding when elements of 1s and 0s are required. Several methods exist to generate \textit{S}-matrices; one of which is to start with a Hadamard matrix, remove the first row and column, and replace 1s with 0s and -1s with 1s. The result is the equivalent of the $H^-$ matrix described above, except with the first row and column removed. An example of an \textit{S}-matrix of rank 3 is given below.

\begin{equation}
    \label{eq:s3}
    S_3 = 
    \begin{bmatrix}
    1 & 0 & 1\\
    0 & 1 & 1\\
    1 & 1 & 0
    \end{bmatrix}
\end{equation}

This paper will focus on HTSI using Hadamard matrix spatial multiplexing as implemented by a dual-channel spectrometer and its performance in the presence of Gaussian and Poisson noise sources.

\section{HTSI simulation using an artificial data set}
\label{sec:simulation}

\subsection{The simulation model}
To simulate HTSI, a simple model was written in the Python language. The model takes in an input data cube, which is then interpolated at specified wavelengths, and each spectral plane is multiplied by a binary mask and then translated left or right according to a dispersion parameter in angstroms/pixel. For example, if the data cube was interpolated at every \textit{m} Angstroms, the dispersion model takes each spectral plane and ``rolls'' it left or right by \textit{m} Angstroms/pixel (a shift of one pixel). This ensures that the output spectrum is neither undersampled nor oversampled. All the shifted spectral planes are then summed together for each mask to produce the multiplexed observations. To simulate the effects of noise, Poisson shot noise is added to the input data, and Gaussian read noise is added after multiplexing. The inverse Hadamard transform is then computed on this output data set and reorganized into a data cube. Two input data sets were used in the simulation: an artificial one-dimensional scene, which is presented in this section, and a spectral data cube of the planetary nebula NGC 7009 (Saturn nebula), which is explored in Section 3.

\begin{figure}
\begin{center}
\includegraphics[width=1\linewidth]{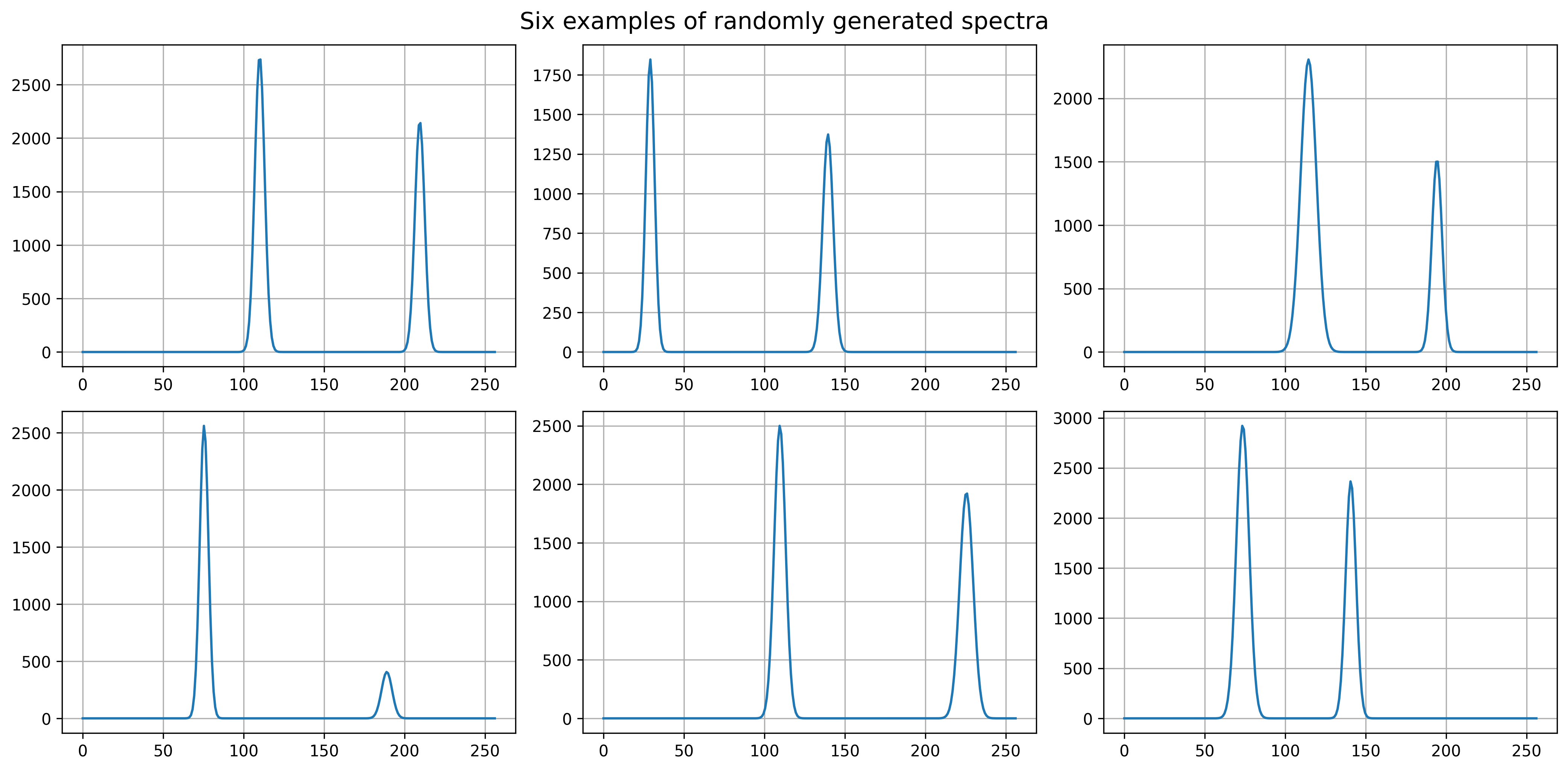}
\end{center}
\caption{\label{fig:spectra_examples} Six examples of artificial spectra are displayed. Each spectrum contains 256 spectral points, and several emission lines with pseudo-random means, widths, and amplitudes. The first line is set to be located in the left half of each spectrum, and the second line in the right half, such that they do not overlap and remain distinct. Axes are in arbitrary units; the \textit{x}-axis is the equivalent to ``counts'' or relative intensity, and the \textit{x}-axis is relative wavelength.}
\end{figure}

The artificial data set contains 64 spatial points and 256 spectral points. Each spectrum contains two randomly generated Gaussian emission lines (random amplitudes, means, and widths) and no continuum (Fig.~\ref{fig:spectra_examples}). This data set can be thought of as a one-dimensional scene that has objects with different spectra in each spatial point. HTSI was computed on this data set using masks generated from a rank 64 Hadamard matrix.

HTSI and single-slit scanning observations were simulated at a range of read noise levels. For slit scanning, the simulation is repeated again with the same data sets, and therefore the same input signal intensity, but instead of the multi-slit masks, single-slit masks are used, which are scanned across the scene (64 positions) and block the light of all but one object. The number of HTSI masks used is equal to that of the single-slit masks. The comparison of the resulting spectra in both HTSI and single-slit scanning cases provides a comparison of multiplexed and direct measurements. HTSI results can also be compared to other direct measurement methods (lenslet arrays, optical fibers, image slicers), however single-slit scanning is chosen as the ``control" method to serve as the basis for SNR gains or reduction achieved by HTSI. It is the most logical and direct method for which to provide a comparison, since both it and HTSI can be implemented on the same instrument using MEMS-based (micro-electromechanical systems) spectrometers with equal slit sizes (discussed in Section 4), and all other instrument parameters will be identical for both cases. In addition, any gains or reduction in SNR with HTSI are inversely proportional to the required observation time to achieve the same SNR in the single-slit case (e.g. an SNR increase of 25\% is equivalent to an observation time reduction of 25\%). Our simulations investigated the improvements achieved with HTSI in two distinct regimes: 1) in the low signal-to-noise regime, where simply the detection of a spectral feature is the goal and 2) in the high signal-to-noise regime, where the parameters of each line, such as line center, line width, etc. are of interest. 

\subsection{Results in the Low SNR Regime}
First, we consider the case of an intrinsically low SNR measurement, where the aim is to observe faint emission lines, against a much fainter continuum. To compare the effectiveness of conventional slit scanning and the HTSI technique, we generated a ``target'' object, with five emission lines with varying degrees of brightness (Poisson-noise limited SNR of 2, 3, 4, 5, and 6). This target spectrum replaced the center spectrum of the scene (index 32 of 64), while all the other spectra were generated in the same manner as in Fig.~\ref{fig:spectra_examples}, with lines having amplitudes randomly sampled from a uniform distribution between 5 and 100, and standard deviations between 0.5 and 2. We then performed simulated observations of the 1D object described in Sec. 2.1, while changing the amount of detector noise (e.g. read noise) present. 

\begin{figure}[h!]
    \centering
    \includegraphics[width=0.8\linewidth]{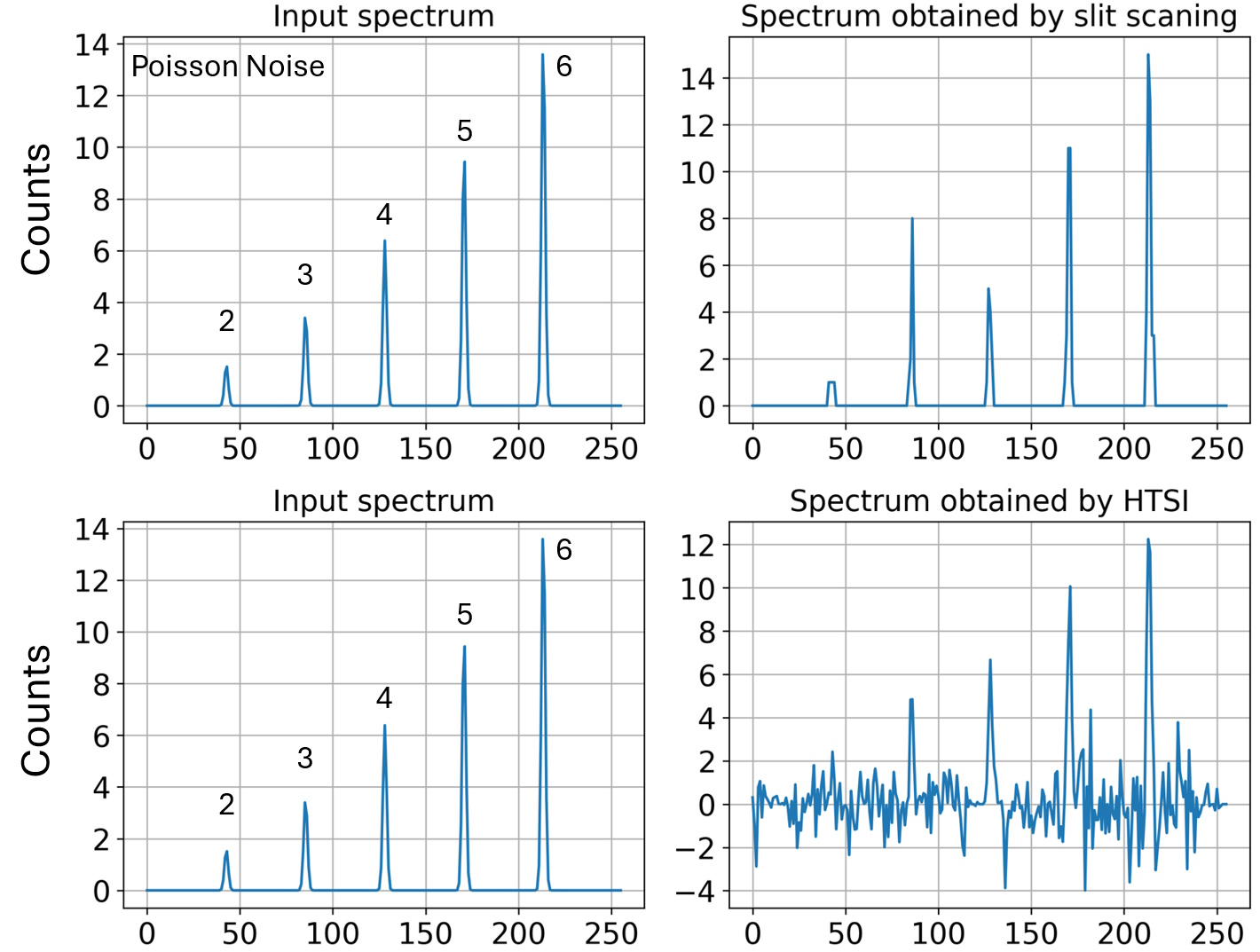}
    \caption{\label{fig:fellgetDisadvantage}An example of a single set of simulated observations of emission lines with intrinsic SNR of 2, 3, 4, 5, and 6. In the case of a perfect detector with zero additional noise, HTSI under-performs when observing the faintest signals, due to the extra noise introduced due to the multiplexing. This is known as Fellget's disadvantage.}
\end{figure}

\paragraph{Fellgett's Disadvantage and Advantage} We first show the results of a perfect detector, with zero noise in Fig.~\ref{fig:fellgetDisadvantage}. Note that while a slit scan with a perfect detector can exactly recover the lines, with no SNR degradation, the HTSI observations show an increased noise floor. This extra noise is due to the multiplexing technique and is known as Fellget's disadvantage. It is a real effect and it ultimately limits the SNR enhancement one might expect from a more simplistic model, which predicts noise suppression on the order of $\sqrt{n}$, where $n$ is the rank of the Hadamard matrix. Nevertheless, when some amount of detector noise is actually introduced, HTSI observations begin to outperform a conventional slit scan, even in the presence of Fellgett's disadvantage. As the detector noise (in this case read noise, but this is true for any signal-independent noise source) increases, the advantage of HTSI becomes clear. Fellgett's advantage (the suppression of detector noise by $\sqrt{n}$) begins to have a greater effect on the SNR, as the multi-slit masks allow more light transmission than a single slit. 

\begin{figure}[h!]
    \centering
    \includegraphics[width=1\linewidth]{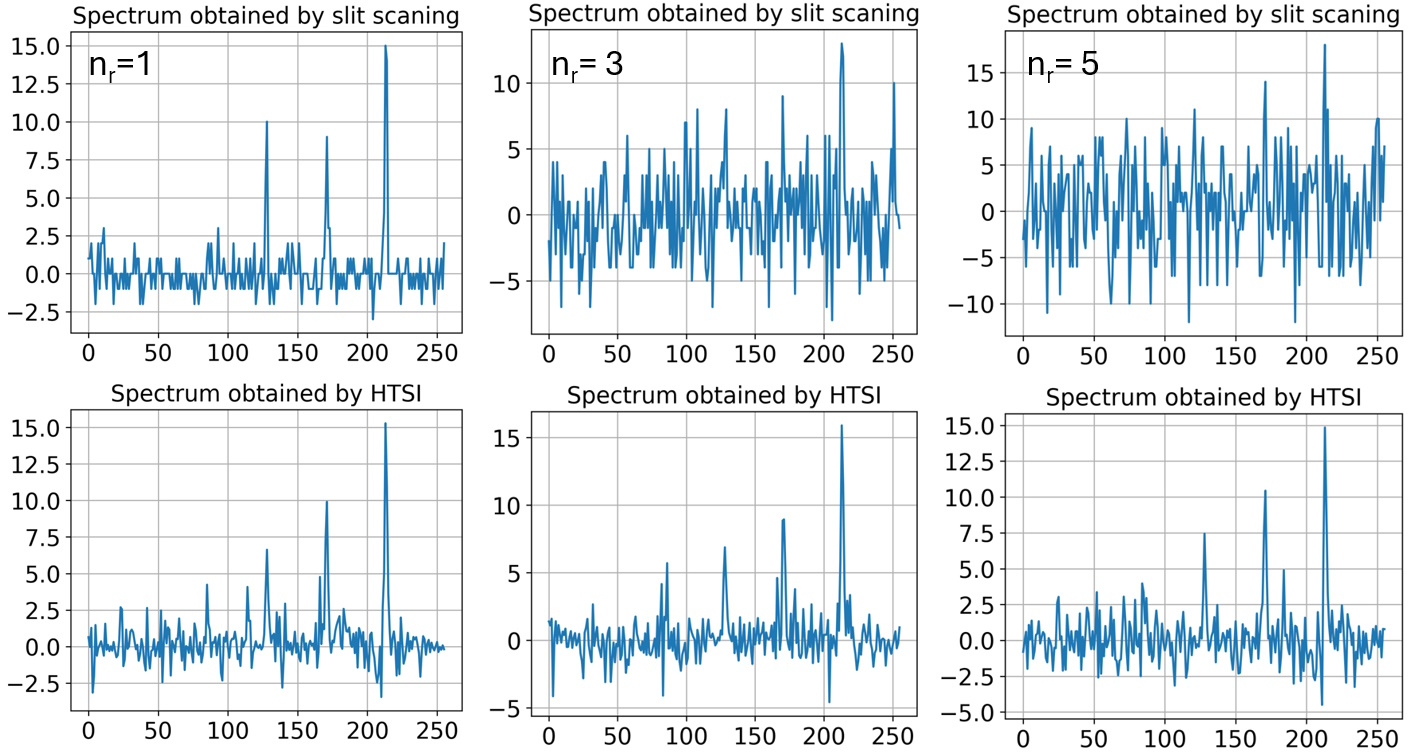}
    \caption{\label{fig:lowSNRLines}Three simulations with the same input spectra as in Fig.~\ref{fig:fellgetDisadvantage}, but with detector read noise of 1, 3, and 5 counts rms. Note that the noise floor in the slit scan measurements increases as one might expect, while the HTSI measurements show a significantly suppressed noise floor - a key goal of multiplexing.} 
\end{figure}

In Fig.~\ref{fig:lowSNRLines} we show three realizations of a simulation with the same input spectrum as in Fig.~\ref{fig:fellgetDisadvantage}, but with read noise of 1, 3, and 5 RMS counts. Note that due to the intrinsically low SNR of the input features, they may not be recovered in any single measurement realization. However, the effect of multiplexing on suppressing the noise floor is very clear - the overall measurement SNR plummets as one would expect for the conventional slit scan, and decreases much more slowly for the HTSI measurements.\cite{Harwit_Sloane_1979,Streeter_Burling-Claridge_Cree_Künnemeyer_2009} This simulation was repeated 50 times, and the average recovered SNRs of the five emission lines across all runs are summarized in Tables 1 and 2 for HTSI and slit scanning, respectively. Table 3 summarizes the measured background standard deviation (noise floor) for HTSI and slit scanning.

\begin{table}[h]
\begin{center}
\caption{The recovered SNRs of the five emission lines with HTSI for various values of detector noise.} 
\bigskip
\begin{tabular}{ c @{\hskip 2cm} c c c c c c c } 
 \toprule
 & \multicolumn{7}{c}{Detector noise RMS} \tabularnewline
 \cmidrule{2-8}
 Poisson SNR lines & 0 & 0.5 & 1 & 1.5 & 2 & 3 & 5 \\
 \midrule
 2 & 1.06 & 1.10 & 1.13 & 0.99 & 1.05 & 1.02 & 0.88 \\

 3 & 2.13 & 2.12 & 2.11 & 2.09 & 2.10 & 2.04 & 1.82 \\

 4 & 3.26 & 3.24 & 3.25 & 3.22 & 3.20 & 3.15 & 2.96 \\

 5 & 4.32 & 4.31 & 4.32 & 4.29 & 4.31 & 4.21 & 4.04 \\

 6 & 5.38 & 5.38 & 5.37 & 5.35 & 5.34 & 5.30 & 5.12 \\
 \bottomrule
\end{tabular}
\end{center}
\label{table:summary_htsi_table}
\end{table}

\begin{table}[h]
\begin{center}
\caption{Similar to Table 1, the recovered SNRs of the lines are shown, but with single-slit scanning.}
\bigskip
\begin{tabular}{ c @{\hskip 2cm} c c c c c c c } 
 \toprule
 & \multicolumn{7}{c}{Detector noise RMS} \tabularnewline
 \cmidrule{2-8}
 Poisson SNR lines & 0 & 0.5 & 1 & 1.5 & 2 & 3 & 5 \\
 \midrule
 2 & 1.97 & 1.50 & 1.08 & 0.81 & 0.66 & 0.47 & 0.37 \\

 3 & 2.91 & 2.69 & 2.25 & 1.82 & 1.47 & 1.09 & 0.77 \\

 4 & 4.00 & 3.72 & 3.32 & 2.85 & 2.49 & 1.84 & 1.19 \\

 5 & 5.02 & 4.79 & 4.41 & 3.99 & 3.51 & 2.82 & 1.87 \\

 6 & 5.89 & 5.78 & 5.43 & 5.01 & 4.57 & 3.63 & 2.56 \\
 \bottomrule
\end{tabular}
\end{center}
\label{table:summary_slit_scanning_table}
\end{table}

\begin{table}[h]
\begin{center}
\caption{The standard deviation of the background.}
\bigskip
\begin{tabular}{ c @{\hskip 2cm} c c c c c c c } 
 \toprule
 & \multicolumn{7}{c}{Detector noise RMS} \tabularnewline
 \cmidrule{2-8}
 Simulation type & 0 & 0.5 & 1 & 1.5 & 2 & 3 & 5 \\
 \midrule
 HTSI & 1.15 & 1.16 & 1.16 & 1.19 & 1.21 & 1.27 & 1.45 \\

 Slit scanning & 0 & 0.57 & 1.04 & 1.53 & 2.01 & 3 & 4.98 \\
 \bottomrule
\end{tabular}
\end{center}
\label{table:background_std}
\end{table}

\subsection{Results in the High SNR Regime}

\begin{figure}[H]
\begin{center}
\includegraphics[width=1\linewidth]{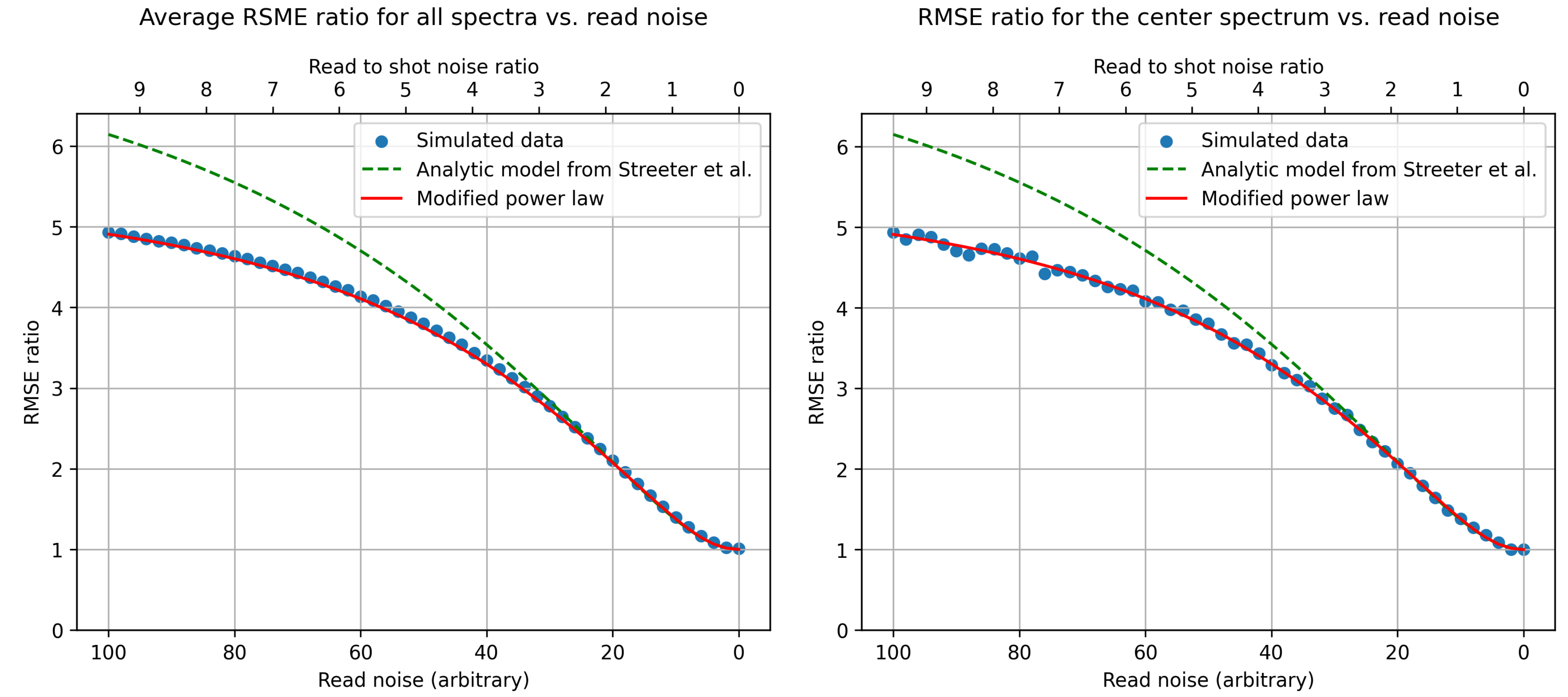}
\end{center}
\caption{ \label{fig:total_RMSE} \textit{Left:} The simulated average RMSE ratio (single-slit to HTSI) for all spectra is shown for various values of read noise (blue dots), along with the theoretical model (dashed green curve) from Streeter et al. (2009). A modified version of this analytic model (red curve) is plotted which better fits the data. For a read noise value of 20, which corresponds to a read-to-shot noise ratio of $\sim$1.9, the average SNR increases by a factor of $\sim$2 with HTSI. At a read noise value of 100 (read-to-shot noise ratio of $\sim$9.4), the average SNR increases by a factor of $\sim$5. When no read noise is present, HTSI has no effect on the average SNR. \textit{Right:} The simulated RMSE ratio for the center spectrum is shown for various values of read noise (blue dots), along with the curves from equations \ref{eq:snr} and \ref{eq:snr_power}.}
\end{figure}

For the high SNR case, the spectra were randomly generated again, but the amplitudes of the lines were scaled by a factor of 30. Thus, the amplitudes were sampled from a uniform distribution between 150 and 3000. In this case, there was no specific ``target" object placed in the data set as in the low SNR regime. To compare HTSI to single-slit performance, several metrics were chosen: the average total root mean square error (RMSE) of all spectra, the total RMSE of a single spectrum, and the standard deviation errors in line means, widths, amplitudes, and line ratios for a single spectrum as obtained from a Gaussian fit of the lines. The fittings were computed on the center spectrum of the scene. Unlike the target object in the previous case, this spectrum does not have pre-defined lines of specific SNR values; it is also randomly generated like the others, with two emission lines. Selecting the center spectrum ensures the maximum amount of overlap with spectra from neighboring spatial points. The ratio of single-slit to HTSI RMSE is used as an analog to the SNR gain achieved by HTSI over single-slit. Figure \ref{fig:total_RMSE} (Left) shows the average RMSE ratio for the entire scene.

For each value of read noise, the simulation was run 50 times, and all outputs were averaged over all runs. The read noise standard deviations range between 0-100 counts, incrementing this value by 2 for each set of runs. The average intensity of all spectra had a value of 112.8 counts, which remained fixed. Thus, the average Poisson noise is $\sqrt{112.8}$ or 10.6 counts, and the read-to-shot noise ratio ranges between 0 and $\sim$9.4. 

The following equation is the theoretical model presented by Steeter et al. (2009) \cite{Streeter_Burling-Claridge_Cree_Künnemeyer_2009} for the SNR gain for HTSI over single-slit, where \textit{$\langle r \rangle$} is the average light intensity over the observation period, \textit{n} is the Hadamard matrix rank (in this case, 64) and \textit{$\sigma$} is the read noise standard deviation:

\begin{equation}
\label{eq:snr}
\text{SNR gain} = \sqrt{\frac{n(\langle r \rangle + \sigma^2)}{n \langle r \rangle+\sigma^2} }
\end{equation}

\noindent The red curve in the figure above is a modification of equation \ref{eq:snr}, with $\alpha=2.41$ and $\beta=2.13$.

\begin{equation}
\label{eq:snr_power}
\text{SNR gain power law} = \left( \frac{n(\langle r \rangle + \sigma^\beta)}{n \langle r \rangle+\sigma^\beta}\right)^\alpha
\end{equation}

The RMSE ratio for only the spectrum at the central spatial point is presented in Figure \ref{fig:total_RMSE} \textit{Right}. Figure \ref{fig:total_RMSE} \textit{Left} shows the average RMSE ratio for all spectra in the scene, however it is useful to select a single spectrum and compare the RMSE ratio to the theoretical models. The simulated data mostly agrees with the model from equation \ref{eq:snr_power}, although there is some slight variation, which is to be expected.

\begin{figure} [!h]
\begin{center}
\includegraphics[width=1\linewidth]{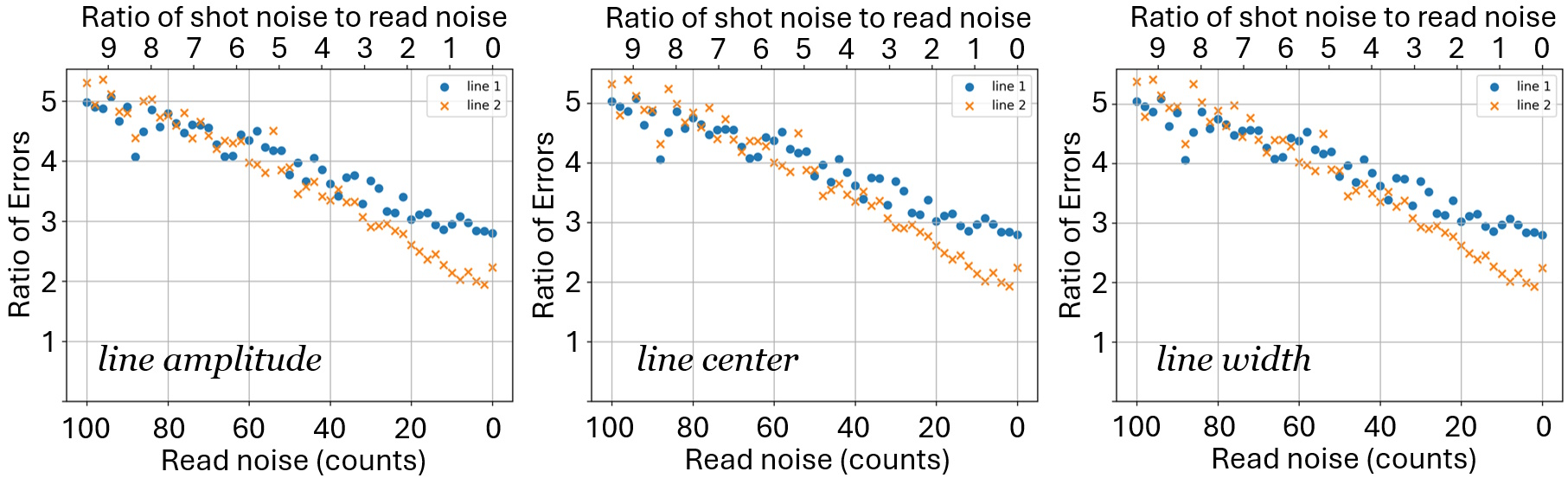}
\end{center}
\caption{ \label{fig:gaussian_fit} The ratio of the standard deviation errors (single-slit to HTSI) for the line amplitudes (left panel), means (center panel), and widths (right panel) as obtained by the Gaussian fit are compared against read noise. }
\end{figure}

Next, the standard deviation errors of the Gaussian fit parameters (amplitude, mean, width) for the two lines in the center spectrum are compared (Figure \ref{fig:gaussian_fit}).

From the presented simulated data, the ratio of the RMSEs (analog to SNR gain) differ from the theoretical model\cite{Streeter_Burling-Claridge_Cree_Künnemeyer_2009}, leading to less of an SNR gain for higher read noise values (a higher read-to-shot noise ratio), but still a considerable gain. For lower values of read noise, the simulated data more closely agrees with the theoretical model. This indicates that HTSI on average performs as well or better than single-slit, however this is highly dependent on the nature of the signal (e.g, whether it contains emission or absorption lines, the continuum level, slope, etc.). Interestingly, the parameters for the lines are recovered more accurately for HTSI than for single-slit observations, even at a read noise level of 0 (when the only noise source is Poisson shot noise), implying that in the high SNR regime, given the same exposure time, HTSI results in higher SNR observations. This is due to the ``error averaging'' effect of the inverse transform: the SNR of the higher signal improves while the SNR of the lower signal degrades, manifesting in the multiplex advantage and disadvantage, respectively \cite{Williams_1989}. In this case, the ``lower" signal is the continuum level (0 photons) and the ``higher'' signals are the emission lines themselves.

\section{HTSI simulation with a 2D object - NGC 7009}
\begin{figure} 
    \includegraphics[width=1\linewidth]{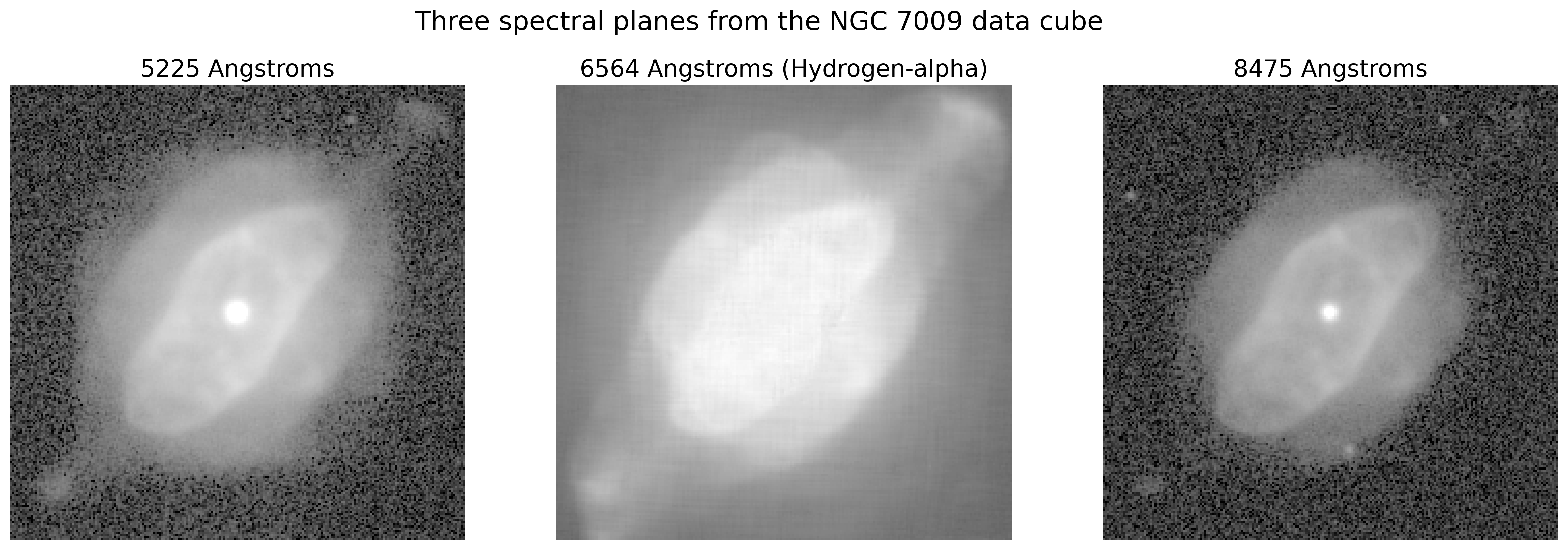}
    \caption{\label{fig:planes}A selection of three spectral planes at various wavelengths from the MUSE NGC 7009 dataset.} 
\end{figure}

The NGC 7009 data cube was obtained from the Multi Unit Spectroscopic Explorer (MUSE) instrument by Walsh et al. (2018)\cite{Walsh_Monreal-Ibero_Barlow_Ueta_Wesson_Zijlstra_Kimeswenger_Leal-Ferreira_Otsuka_2018} and has dimensions of 438 by 441 spatial points and 3802 spectral points. In the simulation, this data cube was interpolated along the spectral axis at every 13 Angstroms, and a subset of 350 spectral planes were selected. For the NGC 7009 input data cube, some examples of the spectral planes are shown in Fig. \ref{fig:planes}.

\begin{figure}[h!]
\centering
   \includegraphics[width=1\linewidth]{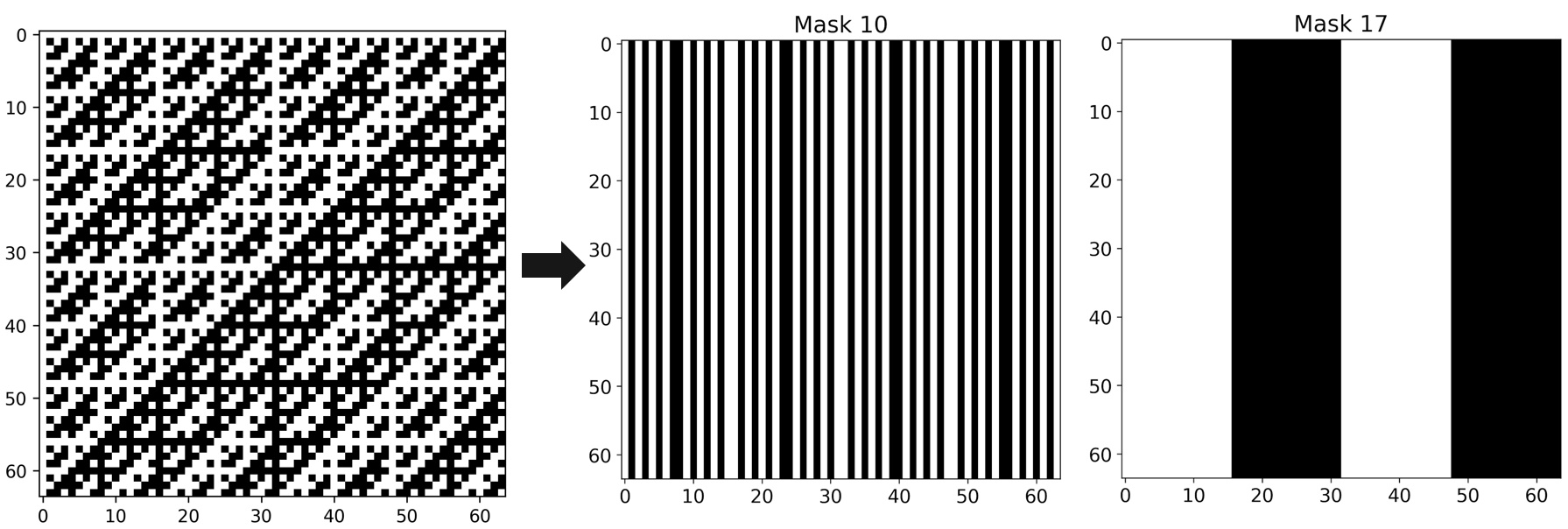} 
    \caption{\label{fig:matrix}A visual representation of a Hadamard matrix of rank 64, with +1 and -1 elements colored as white and black, respectively. The individual binary masks are generated using the elements of a particular row, with examples for row 10 and 17, with only the positive half of the binary pair shown (see Fig.~\ref{fig:elements}).}
\end{figure}

The following figures show the HTSI simulated outputs at various steps in the model to give a better picture of the overall process. First, a Hadamard matrix of order 64 was used to generate the DMD masks. Each row from the rank 64 matrix was used to generate a pair of binary DMD masks (Fig.~\ref{fig:matrix}). Since the Hadamard matrix contains values of  $\pm$1, and the mirrors of a DMD can be turned either ``on'' or ``off'' (1 and 0, respectively), two complementary masks are needed to account for the positive and negative values. As mentioned in the introduction, two masks were generated per row: one for the ``positive'' signal, and the other for the ``negative'' signal. Once the observations are taken with these masks, the data sets are combined into one by subtracting the ``negative'' signal from the “positive” signal \cite{Streeter_Burling-Claridge_Cree_Künnemeyer_2009, Oram_Ninkov_2020} (see Fig.~\ref{fig:elements}).

\begin{figure}[H]
    \centering
    \includegraphics[width=1\linewidth]{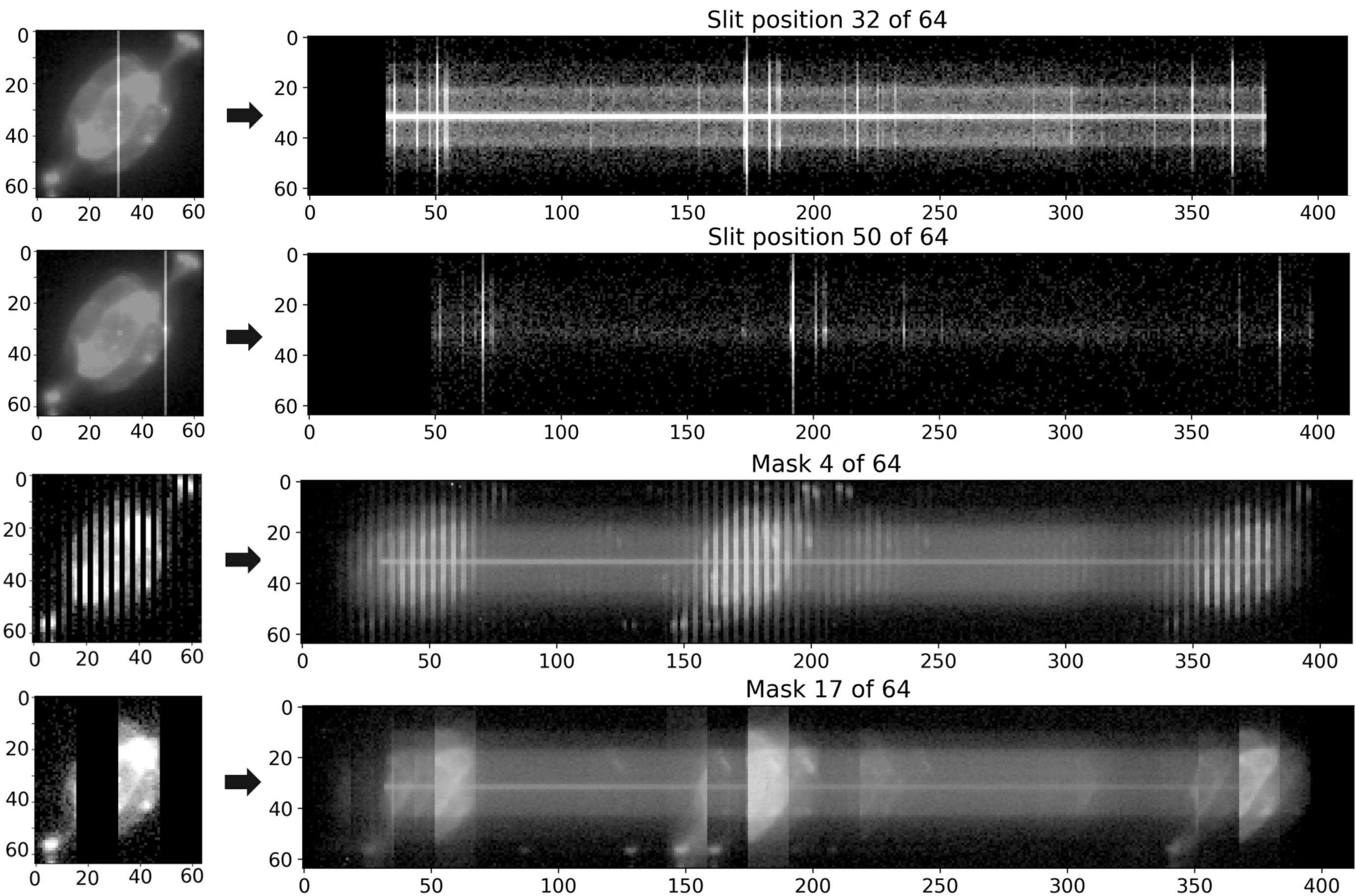}
    \caption{\label{fig:ngc7009Raw} \textit{Top two rows:} Simulated observations of the spectra of NGC 7009 via single-slit scanning. Two examples are shown at different slit positions. The left images show the slit location, while the right panels show the raw spectra. The axes represent pixel indices. In this example, the dispersion direction is along the x-direction, and the right column shows the dispersed light that has passed through the corresponding slit in the left column. \textit{Bottom two rows:} Simulated observations with the Hadamard masks instead of a single slit. These observations are for the ``positive" masks. Due to light passing through multiple slits, there is significant overlap between spectra from neighboring spatial points.} 
\end{figure}

Next, the data cube is multiplied by all the masks and is run through the dispersion model, which simulates the main dispersive element in a spectrometer, such as a grating or a prism. The output of this process results in a one-dimensional convolution of the masks with the spectra along the direction of dispersion. To assess the performance of HTSI later, it is compared to the output of single-slit scanning, similar to the comparison of the artificial data set. Each slit is one DMD mirror wide and is scanned across 64 positions. Figure \ref{fig:ngc7009Raw} show the simulated outputs for single slit-scanning and HTSI obtained for NGC 7009. The inverse Hadamard transform is then computed on this spatially- and spectrally-convolved dataset to recover the spectra, and the resulting data are reorganized into the recovered data cube.

\subsection{Results from the NGC 7009 dataset}

For the NGC 7009 dataset, two spatial points in the recovered output were selected to be compared for single-slit scanning and HTSI. These points are sometimes referred to as “spaxels” (spatial pixels) in the context of a data cube, which are one-pixel slices through the data cube along the spectral dimension. One selected spaxel is located in the center of the nebula, which contains a white dwarf star, and the other is in the upper right stream; the streams are the jet-like protrusions extending outward, which give the resemblance to the planet Saturn, hence the “Saturn nebula”.
   
The simulated output spectra are shown below at the two locations for both single-slit and HTSI with Poisson shot noise as the only noise source. The center location has a much higher photon flux than the stream location, and these spaxels were selected to test the recovery of spectra with different input fluxes.

\begin{figure}[h!]
    \centering
    \includegraphics[width=1\linewidth]{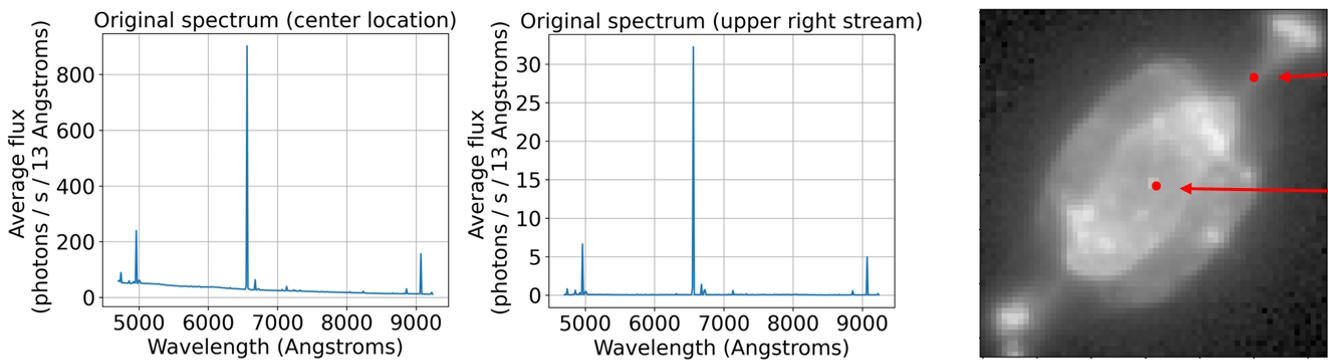}
    \caption{\label{fig:nebulaInputs}The spectra from the two spaxels are shown in the left panel (center point) and middle panel (upper right stream point). The locations of these points are displayed in the right image.}
\end{figure}

\begin{figure}[h!]
    \centering
    \includegraphics[width=1\linewidth]{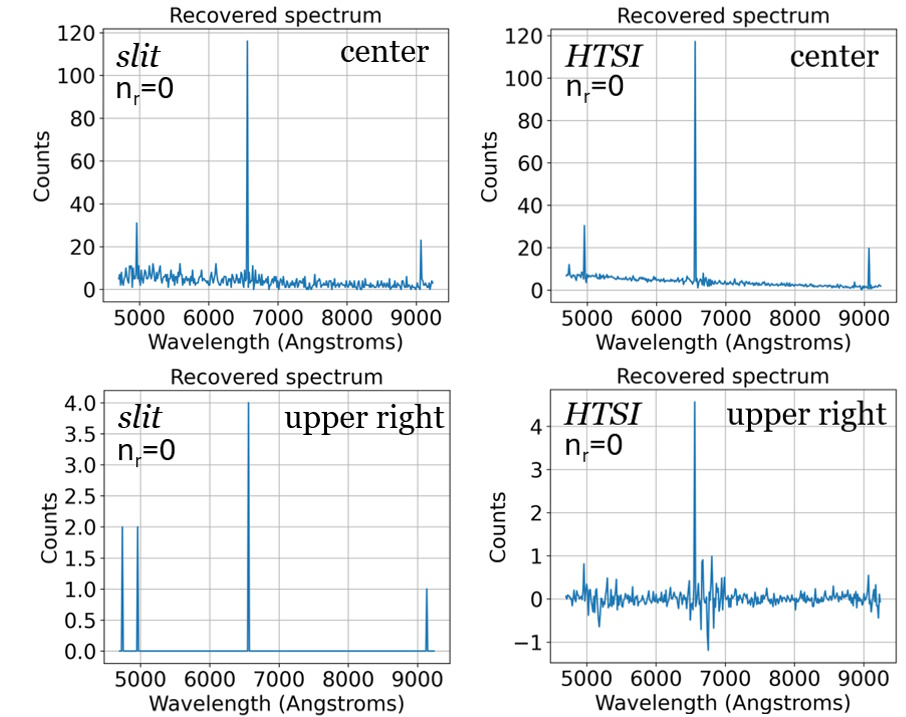}
    \caption{\label{fig:nebulaRN0}A comparison of the recovered spectra at two locations in the planetary nebula from single-slit scanning and HTSI. The left and right columns correspond to slit scanning and HTSI, respectively, and the top and bottom rows show the center and upper right point, respectively. In this example, the detector noise was set to 0.}
\end{figure}

Next, the simulation for NGC 7009 was run again but with a read noise value of 3 electrons.

\begin{figure}[h!]
    \centering
    \includegraphics[width=1\linewidth]{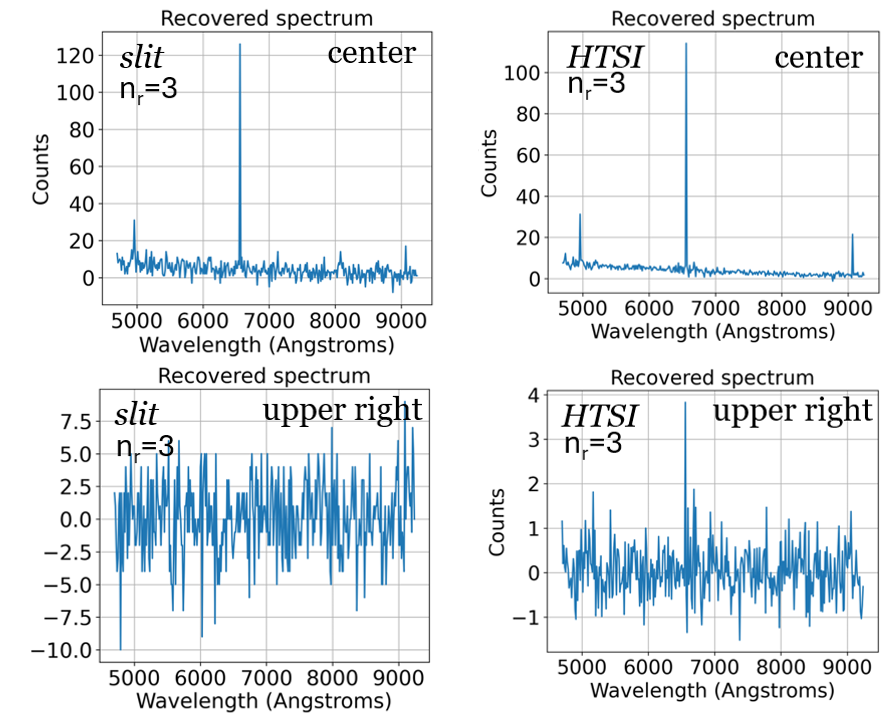}
    \caption{\label{fig:nebulaRN3}Similar to Figure \ref{fig:nebulaRN0}, but with the detector noise set to 3 counts.}
\end{figure}

The results from the NGC 7009 dataset are similar to that from the artificially generated spectra. In the case of no read noise for the point in the stream location, the recovery of the emission lines improve with HTSI, while the continuum (which has a value close to 0 along the entire spectrum) becomes more noisy when comparing it to the single slit observations. Interestingly, for the center point which has a non-zero continuum, all features of the spectrum are improved with HTSI. The average RMSE decreases from 2.01 in the single-slit observations to 0.687 with HTSI, a factor of ~3 increase in SNR.

With a read noise of 3 electrons, the recovery of all features of the spectrum improve with HTSI, similar to the artificial dataset. For the spectrum from the stream location, it is apparent that all emission lines become buried in noise in the single-slit observations. In the HTSI observations, only the bright H-alpha line is recovered in the stream location, which is still an improvement over the single-slit case.

The average RMSE over all spaxels for the single-slit observations are 0.245 and 3.035 for read noise values of 0 and 3 electrons, respectively. For HTSI, the RMSE values for the same two values of read noise are 0.311 and 0.640, respectively. This implies that the recovered data cube from HTSI had average SNR increases by factors of 0.79 and 4.74, respectively. For the case with Poisson noise as the only noise source, the average SNR over the entire data cube decreased with HTSI when compared to single slit observations, and increased in the case with read noise present. An important point to note is that the average RMSE from all spaxels include data with no features that are outside the nebula, primary at the edges of the scene. These spaxels contain very little photon flux and mostly consist of noise already present from the MUSE instrument. As these results are output from one run of the simulation for the NGC 7009 dataset, the average data from multiple runs, similar to that from the artificial dataset, would be needed to draw a more conclusive statement regarding HTSI average effect on SNR.

\section{HTSI with Upcoming Astrophysics Missions}
\label{sec:applications}

\begin{figure}
    \centering
    \includegraphics[width=1\linewidth]{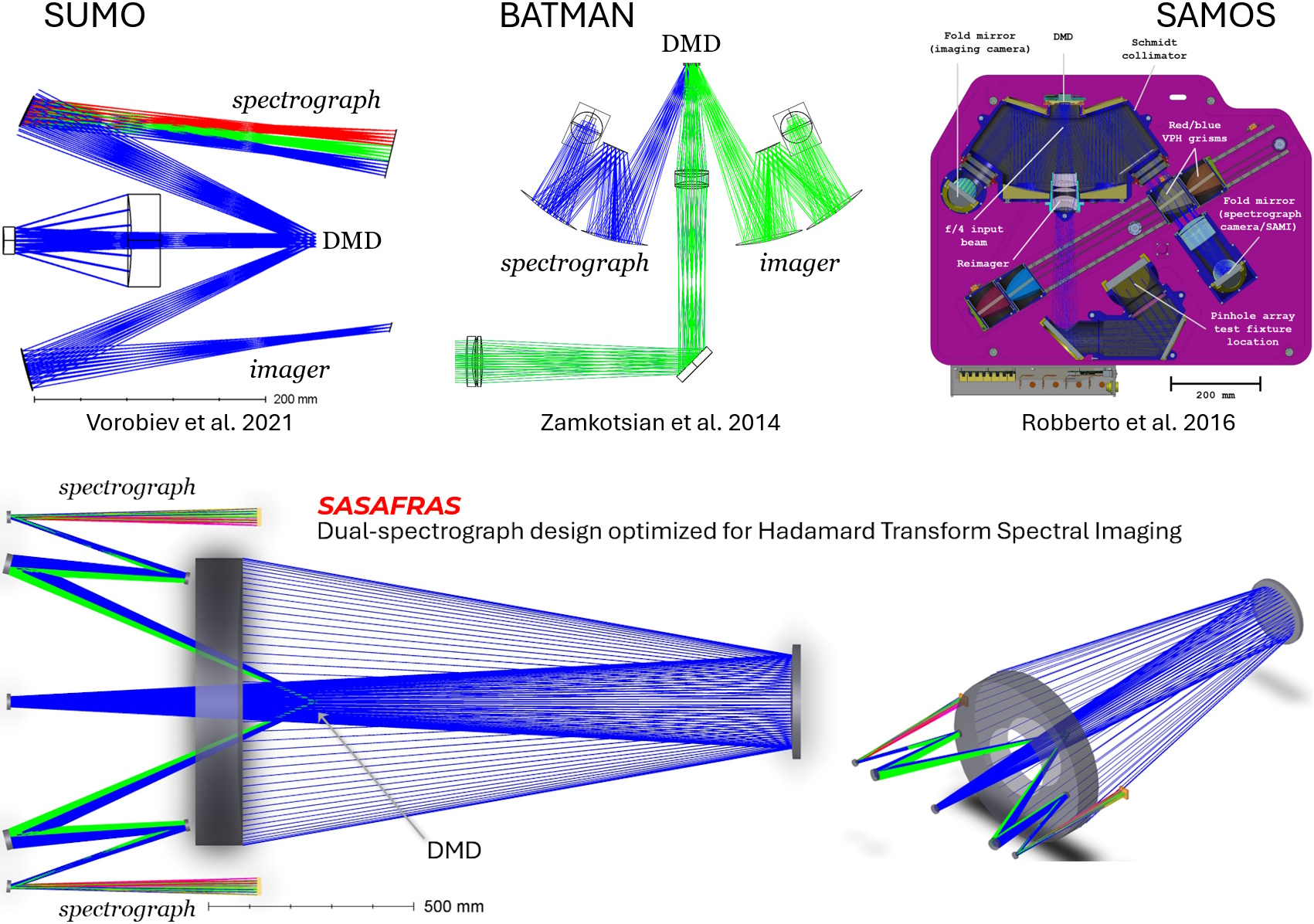}
    \caption{\label{fig:mosDesigns}\textit{Top:} The DMD spectrographs designed and built so far vary significantly, from very simple designs, like SUMO\cite{Vorobiev2021,Halferty2024}, to completely symmetric designs like BATMAN\cite{Zamkotsian2014}, and very sophisticated (SAMOS\cite{Robberto_Donahue_Ninkov_Smee_Barkhouser_Gennaro_Tokovinin_2016}). However, each of these designs have parallel imaging and spectroscopic channels, which may be more suitable for HTSI imaging with $S$ matrices. \textit{Bottom:} Recently, a dual-spectrograph DMD instrument was proposed, designed specifically to benefit from HTSI imaging using Hadamard matrices.}
\end{figure}

The implementation of HTSI can be realized efficiently in practice with MEMS-based spectrographs, with shutters or mirrors to generate the binary masks. As an example, Fixsen et al., (2009)\cite{Fixsen_Greenhouse_MacKenty_Mather_2009} implemented HTSI using a Texas Instruments DMD (digital micromirror array) for infrared spectroscopy. More recently, Robberto et al., developed a DMD-based spectrograph for the visible regime called SAMOS\cite{Robberto_Donahue_Ninkov_Smee_Barkhouser_Gennaro_Tokovinin_2016}. HTSI can also be performed using a microshutter-based spectrograph, such as NIRSPEC\cite{Ferruit2012}  on JWST or the FORTIS spectrograph\cite{McCandliss2017}. Presently, all micromirror and microshutter\cite{Kim2024,Kutyrev2023} based spectrographs have been single channel spectrographs (Fig.~\ref{fig:mosDesigns}), where open shutters/mirrors send light to the spectrograph, while closed shutters simply block the light and ``closed'' mirrors send the light to an imaging channel. To perform HTSI with these instruments, the two binary masks needed for a single HTSI ``observation'' must be performed sequentially. As such, these spectrographs are better suited for HTSI using $S$-matrices. This reduces the number of exposures by half, with the trade-off being a lower SNR gain than the ``proper" HTSI implementation.

Recently, a dual-channel DMD-based spectrograph (SASAFRAS) has been proposed, which can take full advantage of HTSI multiplexing, by using two identical spectrograph channels and two detectors, to obtain the full binary set of observations simultaneously (Fig.~\ref{fig:mosDesigns}).  

One of the key advantages of a MEMS-based (micromirror or microshutter) spectrograph is their inherent flexibility. In addition to the conventional ``slitlet'' multi-object spectroscopy observations, these instruments can also synthesize long slits of varying widths and perform integral-field spectroscopy using HTSI techniques, with either Hadamard or cyclic $S$-matrices\cite{Oram2022}. Multi-object spectrographs (MOS) and integral-field spectroscopy (IFS) are enabling technologies for a wide range of science objectives, with significant ground-based resources dedicated to these instruments. 

For space-based missions, the MOS and IFS toolkit is severely reduced, as compared to ground-based observatories, with the NIRSPEC\cite{Ferruit2012} spectrograph on JWST using microshutter arrays and the FORTIS\cite{McCandliss2017} sounding rocket using a Next Generation Microshutter Array (NGMSA)\cite{Kutyrev2023}. Recently, an integral field spectrograph for the far-UV regime, based on an image slicer, was demonstrated in space for the first time, on the INFUSE sounding rocket\cite{Witt2023}. In October 2025, the SUMO\cite{Halferty2024} Prototype will be the first DMD-based spectrograph to operate in the space environment, as secondary payload on the INFUSE sounding rocket.

\begin{figure}
    \centering
    \includegraphics[width=1\linewidth]{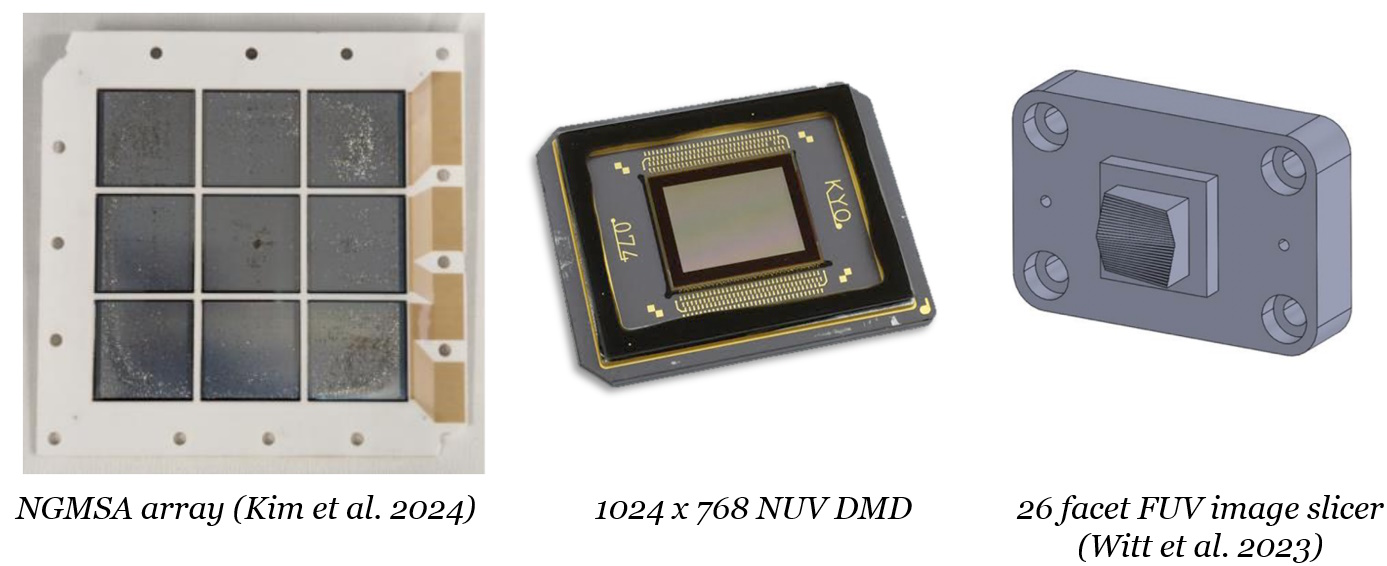}
    \caption{\label{fig:mosTech}Current toolkit for space-based MOS and IFS observations: Next Generation MicroShutter Arrays (NGMSAs), digital micromirror devices (DMDs), and image slicers; shown roughly to scale.}
\end{figure}

As such, microshutter arrays, micromirror arrays, and image slicers (Fig.~\ref{fig:mosTech}) are currently the most mature technologies for MOS and IFS observations in space. Several projects, which span the range of mission scale, are currently in development, with multi-object spectroscopy as part of their baseline capability. The smallest of these is the Supernova And Stellar Feedback Rapid Acquisition Spectrograph (SASAFRAS) (Fig.~\ref{fig:mosDesigns}), a dual-channel DMD spectrograph optimized for Hadamard matrix multiplexed spectroscopy on a NASA long duration super-pressure stratospheric balloon.  

The Cosmological Advanced Survey Telescope for Optical and ultraviolet Research (CASTOR)\cite{Cote2024} is a 1.5~m telescope for wide field and high angular resolution imaging and spectroscopy, currently in development. The baseline CASTOR mission includes multi-object spectroscopy using a slitless observations with a grism. However, the CASTOR team is actively investigating the possibility of including a DMD-based multi-object spectrograph, in the traditional (single channel, Fig.~\ref{fig:mosDesigns}) configuration. The CASTOR UVMOS would be capable of HTSI observations, significantly expanding its capability, without requiring any extra hardware.

Lastly, and most significantly, multi-object spectroscopy is a baseline capability of the Habitable Worlds Observatory, NASA's upcoming flagship telescope. Although the capabilities of HWO are currently being established, it is likely that HWO will have MOS capability using a microshutter array (or a DMD) and integral field spectroscopy with an image slicer instrument. The IFS observations enabled by a MSA or DMD MOS are complimentary to those obtained with a slicer-based IFU. Specifically, the current slicer IFU design aims for a field of view of $\sim3\times3$ arcseconds, with spatial resolution of $\sim100$ mas. The MSA MOS would be capable of IFS observations over a much larger $\sim2.6\times2.6$ arcminute FoV, with tunable spatial resolution (with smallest possible single-shutter sampling of $0.2\times0.4$ arcseconds). These HTSI observations will provide lower spatial and spectral resolution than the slicer IFS, especially over a small field of view. However, the ability to spectrally survey large areas of the sky may be extremely useful for observations of very faint, extended objects.

\section{Conclusion and Future Work}

The results from the simulated observations of single-slit scanning and HTSI show that HTSI improves the SNR of spectra in the presence of Gaussian read noise, and this effect is most noticeable on emission lines when the signal level is low. As the ratio of Poisson shot noise to Gaussian read noise increases, the average SNR gain achieved by this technique gradually decreases, converging asymptotically to a value of unity when no read noise is present. This is confirmed by the results from the artificial data set, which seem to agree with Streeter et al.\cite{Streeter_Burling-Claridge_Cree_Künnemeyer_2009}. Although the SNR gain for higher read noise values is slightly less than predicted by the analytic model\cite{Streeter_Burling-Claridge_Cree_Künnemeyer_2009}, HTSI still achieves a considerable SNR advantage over single-slit scanning. 

Results from simulated observations of the NGC 7009 data cube are mostly similar to that of the artificially generated spectra. However, as this dataset has varying continuum fluxes across all spaxels unlike the artificial dataset, and some spaxels that were included in the average RMSE calculation are outside of the main nebula, a conclusive statement cannot be made unless the average results of multiple runs of the simulation are taken into account due to randomness of Poisson and Gaussian noise sources. Unlike the results for the artificial dataset, which averaged the outputs from 50 runs, the results for the NGC 7009 data cube were obtained from only one run of the simulation. The output can vary from one run to another due to the randomness of the noise, therefore an average of multiple runs is needed for a more accurate statement, although the results are expected to converge to that of the artificial data set with multiple runs.

Future work will include further analysis of the results from the NGC 7009 dataset as stated above, as well as simulating spectra with both absorption and emission lines, and testing how these features are affected by HTSI under various noise sources. In addition, the variability of the photon flux during the observation time with HTSI is another effect to be modeled in the future. For example, variable atmospheric conditions, such as obstruction by clouds, during an observation with one mask but not others will affect the entire collected dataset in HTSI, but not for single-slit scanning. With single-slit scanning, data would be missing for one slit position, but with HTSI, the data would still be included in half of the other observations. When the inverse transform is computed on a dataset with a missing observation from one mask, an estimate of the original spectra is then recovered. Investigating methods for recovering missing data with HTSI is also planned, as this could provide another advantage over single-slit scanning.

\section{Disclosures}
The authors declare that there are no financial interests, commercial affiliations, or other potential conflicts of interest that could have influenced the objectivity of this research or the writing of this paper.

\section{Code and Data Availabiltiy}
Data availability statement: although there is no data to be made available in this work, the authors would be open to share the code used to perform the computations.

\acknowledgments 
 
The authors would like to acknowledge Stavros Akras \cite{Akras_Monteiro_Walsh_García-Rojas_Aleman_Boffin_Boumis_Chiotellis_Corradi_Gonçalves_et_al._2022} for sharing the MUSE data cube of NGC 7009 that was obtained by Walsh et al. (2018) \cite{Walsh_Monreal-Ibero_Barlow_Ueta_Wesson_Zijlstra_Kimeswenger_Leal-Ferreira_Otsuka_2018}. This work was supported by NASA APRA grant 80NSSC21K1669, and is an expansion of the SPIE conference proceeding paper 13383-12.

\bibliography{ref}   
\bibliographystyle{spiejour}   

\end{document}